# Few layers graphene on 6H-SiC(000-1): an STM study

François Varchon, Pierre Mallet, Laurence Magaud, and Jean-Yves Veuillen

Institut Néel, CNRS et Université Joseph Fourier, BP 166, F-38042 Grenoble Cedex 9,
France

We have analyzed by Scanning Tunnelling Microscopy (STM) thin films made of few (3-5) graphene layers grown on the C terminated face of 6H-SiC in order to identify the nature of the azimuthal disorder reported in this material. We observe superstructures which are interpreted as Moiré patterns due to a misorientation angle between consecutive layers. The presence of stacking faults is expected to lead to electronic properties reminiscent of single layer graphene even for multilayer samples. Our results indicate that this apparent electronic decoupling of the layers can show up in STM data.

73.20.-r, 68.55.-a

## I. Introduction

Graphene has received a lot of attention in the last few years due to its very appealing transport properties[1,2,3,4]. Most of the work has concentrated on samples produced by mechanical exfoliation[5] and contacted using lithographical techniques. Apart from mechanical exfoliation, a convenient way to prepare single layer graphene or few layers graphene (FLG) samples is the thermal decomposition of the hexagonal faces of SiC single crystals[6]. It has long been known that heating at high temperature polar faces (either Si or C terminated) of 6H or 4H-SiC substrates in vacuum leads to Si sublimation and to the formation of carbon layers in a graphitic form at the surface[7]. Actually, transport measurements and infra-red measurements have shown the existence of Dirac fermions in such FLG's samples [3,8]. This has generated a lot of activity for the investigation of this material by means of modern surface science techniques. Experiments aim at elucidating the atomic and electronic structure of the system in this favourable situation where electron states that give rise to the fascinating electronic properties of the material can be directly probed by either Angle Resolved Photoemission Spectroscopy[9,10,11] (ARPES) or STM[12,13,14]. It turns out however that up to now most of these studies have been performed on the Si terminated face, although most of the transport measurements have been made on the C terminated face which shows higher carrier mobility [3,15]. It is therefore desirable to gain more information on FLG's formed on this SiC(000-1) surface, known as the C terminated face. We present in this paper an investigation of the morphology and atomic structure of FLG's on 6H-SiC(000-1) by means of Scanning Tunneling Microscopy (STM).

The observation by low energy electron diffraction (LEED) of diffraction rings for FLG's grown on 6H(4H)-SiC(000-1) indicates a significant amount of azimuthal disorder in the films [7,16,17,18,19]. A recent structural investigation by X-ray reflectivity on relatively thick FLG's (4-13 graphene layers) grown on 4H-SiC in an induction furnace indicates that disorder arises from stacking faults in the film, this is from a misorientation between adjacent layers [17]. Since the stacking faults alter the AB (Bernal) stacking of crystalline graphite, this would have a strong influence on the electronic structure of the layers [15,17]. Indeed, recent theoretical calculations have shown that the electronic structure of misoriented bilayer (or trilayer) graphene is quite different from the one of AB stacked layers. For either large[20,21] or small rotation angle[22] a linear (Dirac like) band dispersion is recovered close to the K point, whereas AB stacked bilayers show a parabolic behaviour. In this scheme, the presence of stacking faults in FLG's formed on the C face would explain the unexpected occurrence of graphene (single layer) properties in multilayer samples [20,21]. Investigating the nature of the disorder is thus an important issue for FLG's samples. STM experiments, which offer a local view of disorder in real space, nicely complement (more integrating) diffraction techniques for that purpose. We report here the observation of a significant amount of stacking faults with various rotation angles (including small ones) between adjacent layers for thin layers (3-5 graphene planes) grown under UHV conditions on 6H-SiC. This work provides a direct evidence for the rotational disorder in the FLG's grown on the C face of hexagonal SiC polytypes.

## II. Experiment

The sample preparation procedure is similar to the one reported before [7,14,18]. The surface of the 6H-SiC(000-1) sample (n-doped, purchased from NovaSiC) surface was first cleaned under ultra-high vacuum (UHV) by heating at 850° C under a silicon flux. After annealing at 950-1000°C, the surface showed a (3x3) reconstruction similar to the one reported[23] from LEED and Auger Electron Spectroscopy (AES). FLG's were formed on this surface by further heating for 15 minutes at temperatures close to 1150°C, where multilayer growth has been reported [7]. Two different annealing temperatures were used, which were slightly below and above 1150°C (within 50°C: the accurate determination of the temperature with the pyrometer is difficult since the sample is transparent). The thickness of the FLG's can be estimated from AES to be 3±0.5 and about 5±0.5 graphene layers respectively (notice that no signal from the underlying interface could be detected by STM, which indicates that even the thinnest sample was more than 2 layers thick by comparison with the Si face [14]). In both cases the FLG's exhibit a ring-like LEED pattern, with more intense spots (reinforced intensity) along specific substrate directions, as already reported [7 16 18 19]. This is indicative of azimuthally (but not randomly) disordered films. The STM experiments were performed at room temperature in UHV using mechanically cut PtIr tips. The STM observations reported in this paper were similar for the two layers (3 and 5 ML thick).

In the whole paper, AB refers to the stacking sequence of carbon planes and $\alpha$ and $\beta$ refer to the two sites in the unit cell of the surface graphene layer. For bulk Bernal graphite for instance, the stacking is …ABAB…, the $\alpha$ site is above a carbon atom in the next layer, whereas the $\beta$ site is on a hollow site (they are therefore not equivalent).

## III. Results and discussion

A representative large scale (150x150 nm²) image is shown in figure 1. In figure 1-a, there is essentially a single terrace (see below) cut by pleats (P) with typical height 0.5-2 nm. Such P features have already be mentioned for graphitized 6H-SiC(000-1) surface, although for a much higher annealing temperature[24]. One also notices curved lines made of "beads" (B), with typical height 0.2 nm, which were also observed in a previous work [19]. Atomic resolution of the P and B structures demonstrate that they are made of (curved) graphitic carbon (a small scale image of a B structure is shown in figure 3-a), as in Ref. 24. This kind of features is generally not observed for FLG's grown on the Si face [14,25]. Their origins are unknown, but they have been considered as precursors for the growth of carbon nanotubes on the C face [19]. Figure 1-b is the same image as figure 1-a, but with an enhanced contrast on the flat area. The difference in height on the coloured (light grey) area is less than 0.2 nm (FWFM), which shows that it is a single terrace. Atomic resolution images taken at various spots on the flat areas (see e. g. figure 2-a) reveal a hexagonal structure with the lattice parameter of graphite (a=0.246 nm), as expected. An important point is that various superstructures (super lattices) are observed on the terraces, which are bounded by B or P structures. Their period is in the nanometre range (from 2.5 to 3.8 nm in figure 1-a), and their corrugation is a fraction of Å. They resemble the superstructures which have been extensively studied on graphite[26,27,28], and which are interpreted as moiré pattern due to a misorientation, with rotation angle $\theta$, between the two outermost C layers[29]. In the following, we present arguments which support this interpretation, but we already notice that the observation of these "moiré patterns" is a direct evidence for a rotational disorder in the (vertical) stacking of the FLG's.
The interpretation of the superstructures in figure 1-a as moiré patterns is made using the same arguments as for graphite surfaces [27,28,29]. In figure 2-a, we show an enlarged view of the boxed area of figure 1-b, around the boundary between the flat and corrugated zones. Atomic resolution is achieved on the whole image, and the Fourier Transform (FT) of the image (figure 2-b) shows that the atomic lattice of the surface graphene layer rotates across the boundary. The rotation angle is close to 5°. The period of the superlattice on the right side of the image is D=2.8 nm. In a moiré picture [27 28 29], assuming that the underlying C plane has a unique orientation, one expects $D=a/(2\sin(\theta/2))$, this is D=2.82 nm for θ=5° and a=0.246 nm, which is consistent with the measured value. One can also measure the angle between the main axes of the superstructure and of the surface atomic lattice, shown as φ in figure 2-c (on another spot of the sample). In the moiré picture [27 28 29], θ and φ are related by φ=30°-(θ/2). From the measured value of D (1.5 nm) we derive θ=9.44° and we expect φ=25.3°, in agreement with the measured value of 25±2° (the measured value of this angle is affected by the STM drift). The consistency between D, θ and φ has been verified on a number of different superlattices, which definitively establishes the origin of these structures as moiré patterns. Other features such as the presence of "beads" B and the typical corrugation of the superstructure (0.2-0.5 Å) are also reminiscent of the "moiré patterns" observed on graphite surface [27 28 29]. Notice that in some cases (as in figure 3-a) the moiré pattern is found without any rotation in the surface lattice, which indicates a change in the orientation of the underneath layer at the boundary.

We have observed superlattices with period D ranging from less than 1 nm up to 10 nanometers. These values of D correspond to rotation angles θ between adjacent planes ranging from 1.5° to 19°. In agreement with recent computations[30], small period (around 1 nm) superlattices are difficult to detect in large scale images, not only due to their small wavelength but also due to their reduced corrugation. There are also areas without "moiré patterns" (e.g. figure 1-b, upper left), which therefore correspond to normal AB stacking at the surface. Atomic resolution shows the usual triangular contrast of graphite due to AB (Bernal) stacking in this area (notice however that even these regions were not absolutely flat, showing long range modulations with amplitude of a fraction of Å). Therefore there is a wide distribution of stacking angles in the samples.

In figure 1-b, one notices that the orientation of the superstructures is different by approximately 20°, in the lower an in the upper part of the figure, although the period D is roughly the same. This is due to a similarly large ($\cong 20°$) rotation of the atomic lattices across the pleat, which is observed in atomic resolution images. For these large period superstructures (D ≥2.5 nm) φ is close to 30° (within less than 3°). The rotation of the superlattice therefore follows the rotation of the atomic lattice. This is a convenient way to identify different orientations of the surface atomic lattices between adjacent grains on the surface in large scale images ("grains" here refer to areas separated by P structures). To complete the characterisation of the rotational disorder, we report two additional characteristics. Firstly, we have observed directly at some spots a rotation of the atomic lattice in the surface layer by large angles (20-30°) without any P structure, one example being shown in figure 2 d and e. In figure 2 d (and also in figure 2-a), in addition to the atomic lattice, one can see the well known √3x√3R(30°) (or R3) superstructures [12][14] which are due to electron scattering at the boundary between regions I and II (their directions are indicated by dashed lines). Secondly, super-lattices with two periodicities were found to coexist in some areas (not shown). Although the origin of this phenomenon is not established [29], it may indicate a stacking of three graphene planes with different rotation angles.

The picture which emerges from this structural study is that there is a significant azimuthal disorder in the FGL's grown on the C face, both between the grains and inside the grains (stacking disorder). Although the former type of disorder is certainly detrimental for the properties of the material, the later may help restoring the electronic properties of single layer graphene even in multilayer samples, as mentioned in the introduction. It is interesting to consider the effect of this apparent electronic decoupling [20][21] of rotated layers on the STM images of graphene. Naïvely we could expect to recover the honeycomb contrast of isolated single layer graphene where all atoms ($\alpha$ and $\beta$ type[31]) show up in STM data [12][13][14], at variance with the $\alpha/\beta$ asymmetry found for Bernal AB stacking which leads to a triangular contrast [12][14][31]. Actually, STM images computed for trilayer samples show that it may be the case [20]. Notice however that i) the result depends on the stacking order and on the sample bias for a given stacking and ii) that the computation have been made for large misorientation angles (16°) compared to the ones we usualy observe. Although the linear dispersion has also been predicted for lower angles [22], no simulation of STM images has been made. It turns out that we frequently observe a seemingly honeycomb pattern (or very weak $\alpha/\beta$ asymmetry) on small period lattices (see e. g. figure 2-c). To verify that such contrast is not due to tip artefacts, we have chosen to image boundaries between "flat" regions without superstructures (and thus presumably normal Bernal AB stacking) and regions with a superlattice. In this way, we get a reference for the tip on the (flat) region of AB stacked layers with triangular contrast[32]. We observe in general a significantly reduced $\alpha/\beta$ site asymmetry on the superstructure compared to the flat region. This is the case for instance in figure 2-a: although the contrast remains essentially triangular on the whole image, the $\alpha/\beta$ site asymmetry is significantly smaller on the right side of the boundary. This has been reported previously on a large period (6.6 nm) pattern on graphite [28], and this is not unexpected considering that the strict Bernal A/B stacking is lost over most of the superlattice cell [27][28][29][30] (in particular, the stacking is supposed to be close to AA in the vicinity of the brightest points [28][30]). In some cases, the asymmetry is reduced to the point that we observe a contrast similar to the one of isolated graphene layers (i. e. $\alpha$ and $\beta$ sites appear with the same contrast [14]) in a range of small positive and negative biases, this is for energies that straddle the Fermi level. One example is shown in figure 3. The upper part (figure 3-a) is a view of the boundary, showing B structures with atomic resolution and a superlattice on the right side (D=2.25 nm). Images 3-b to e are extracted from similar images at lower bias (+200mV for Figures 3-b and 3-d, -200mV for 3-c and 3-e) on the two sides of the boundary (left for figure 3-b and 3-c, right for figure 3-d and 3-e). It is clear that the contrast is different on the right and left side of the image, with a vanishingly small –if any- $\alpha/\beta$ site asymmetry on the superlattice and a clear triangular contrast (strong $\alpha/\beta$ asymmetry) on the flat region. We do not claim that this is a direct or unambiguous proof of the electronic decoupling of the layers since i) the calculations of Reference 20 suggest that a triangular contrast may show up even in the presence of graphene-like dispersion depending on the stacking and on the bias (and we do not have access to the whole stacking sequence), and ii) a AA stacked bilayer should show the same honeycomb contrast although the layers show a significant interaction[33]. We consider however that it is a valuable indication that a single layer-like behaviour can be found on rotated layers.

## IV. Conclusion

To conclude, we have investigated by STM the morphology and the atomic structure of FLG's (3-5 layers) grown on the 6H-SiC(000-1) surface by graphitisation under UHV. Our real space observations reveal a significant amount of rotational disorder the films, in agreement with previous structural studies. The surface present Moiré patterns which directly demonstrate misorientations in the stacking of the planes in addition to an azimuthal disorder between grains. The rotational disorder has been shown to affect the electronic properties of the FLG's, and our experimental data suggest that these changes can be observed by STM.

## Acknowledgments

This work is supported by the ANR (projet "GraphSiC") and by the Program "Cible07" of the Région Rhône-Alpes. We acknowledge D. Mayou, C. Naud and F. Hiebel for fruitful discussions.

**FIG. 1: (Color online) (a) Large scale image (150x150 nm²) of a terrace for the graphitized 6H-SiC(000-1) surface. Some pleats (P) and beads (B) structures are indicated. Sample bias: $V_s$= +1.0V, tunnelling current $I_t$=0.1 nA. (b) Same image as in (a) but with an enhanced contrast on the flat area. Superlattices with periods in the nm range are seen on the terrace, bounded by P or B structures. Notice the**

**different orientation of the superstructures in the upper and lower part of the image.**

**FIG. 2: (Color online) (a) Image of the boxed area of figure 1-b. Image size: 20x20 nm², $V_s$= -0.25 V, $I_t$=0.1 nA. The dotted (dashed) line underlines the direction of the atomic rows in the left (right) part of the image; with a relative angle of 5°. (b) Fourier transform of the image in figure 2-a. The outer spots marked (unmarked) by the arrows correspond to the atomic lattice on the right (left) side of the image. Their relative orientations are again rotated by 5°. (c) 12x7 nm² image of a superlattice with period D=1.50 nm, $V_s$= -0.5 V, $I_t$=0.3 nA. The dotted line gives one direction of the superlattice, and the dotted-dashed line one direction of the atomic lattice. φ is the angle between these directions. We measure φ=25±2° from several images of this area. (d) Boundary with a large rotation angle between the surface atomic lattices in regions I and II without P or B structure. Dashed lines indicate the directions of the superstructures induced by electron scattering at the boundary. $V_s$= +0.1 V, $I_t$=1.0 nA(e) Fourier transform of image in figure 2-d. Circled (not circled) outer spots correspond to the atomic lattice in region I (II). The inner spots correspond to the R3 superstructure.**

**FIG. 3: (Color online) (a) 15x7 nm² image of a boundary between a flat area (left) and a superstructure (right), $V_s$= +0.5 V, $I_t$=0.4 nA. (b) and (c): Images of the boxed area on the left side of figure 3-a for sample biases $V_s$= +0.2 V and -0.2 V respectively. (d) and (e): Images of the boxed area on the right side of figure 3-a for**

sample biases $V_s$= +0.2 V and -0.2 V respectively. Size of the images in (b) to (e): 4x4 nm². $I_t$=0.2 nA for (b) and (d) and $I_t$=0.4 nA for (c) and (e).

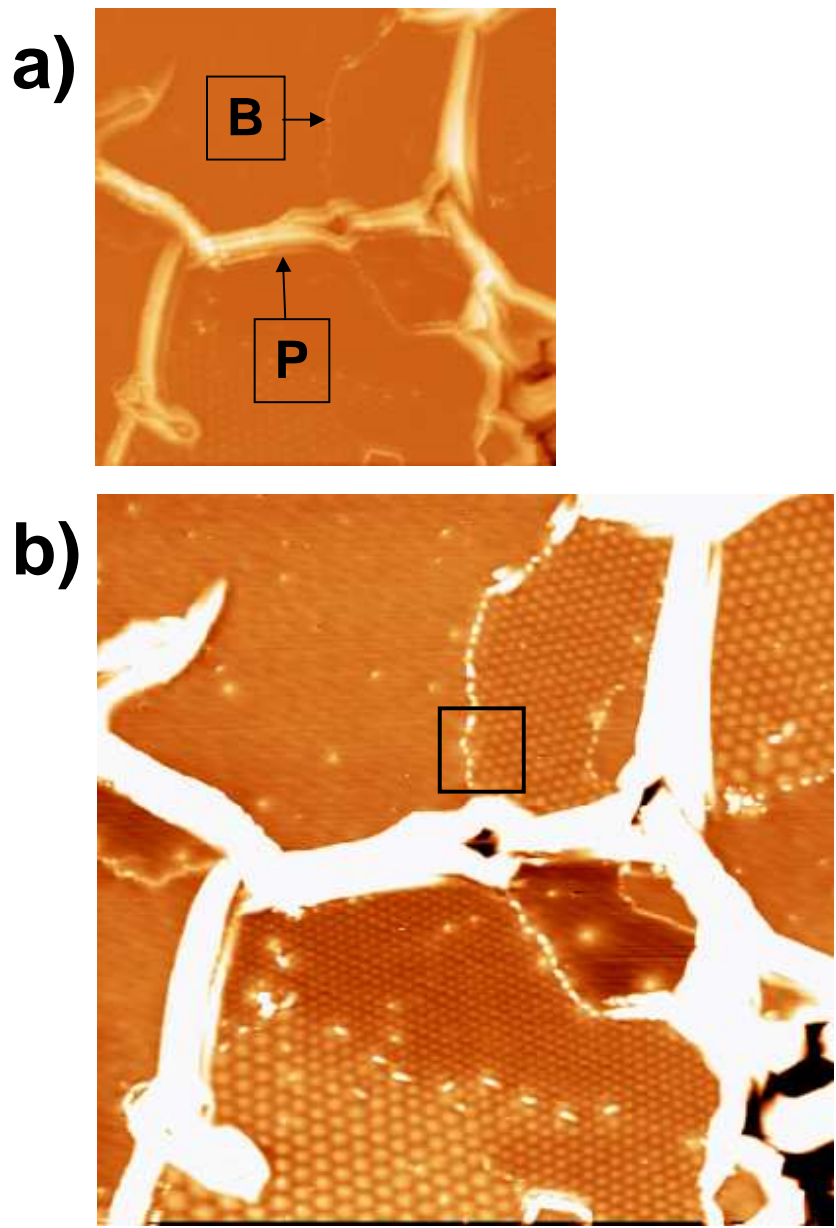

Figure 1

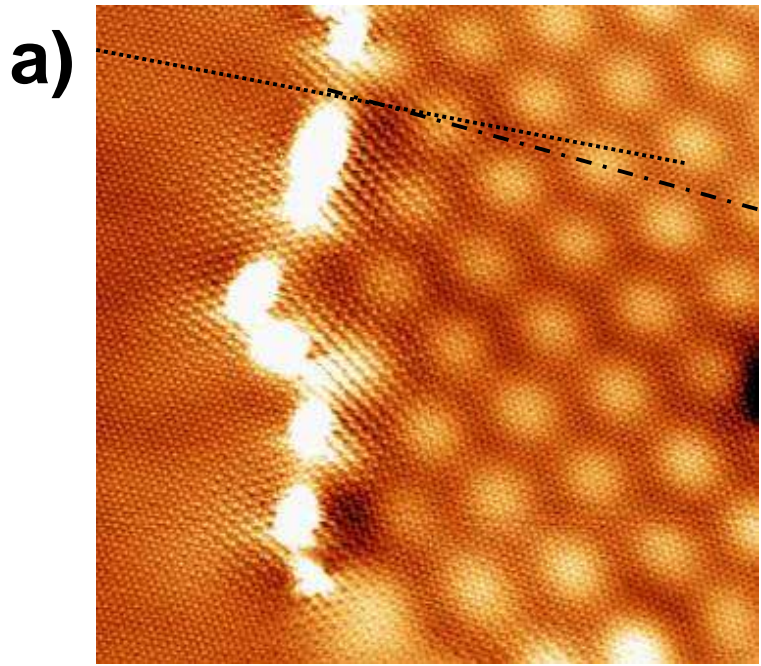

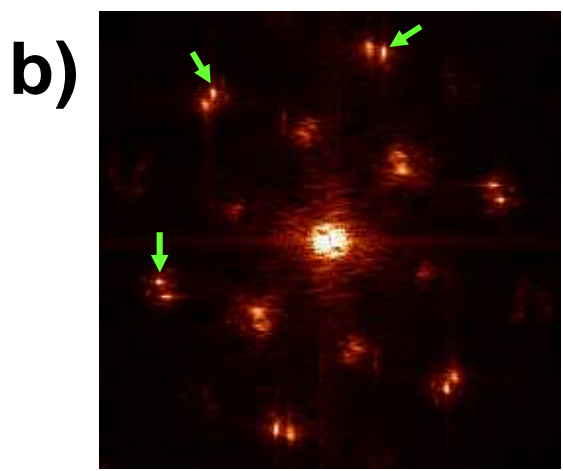

Figure 2 a-b.

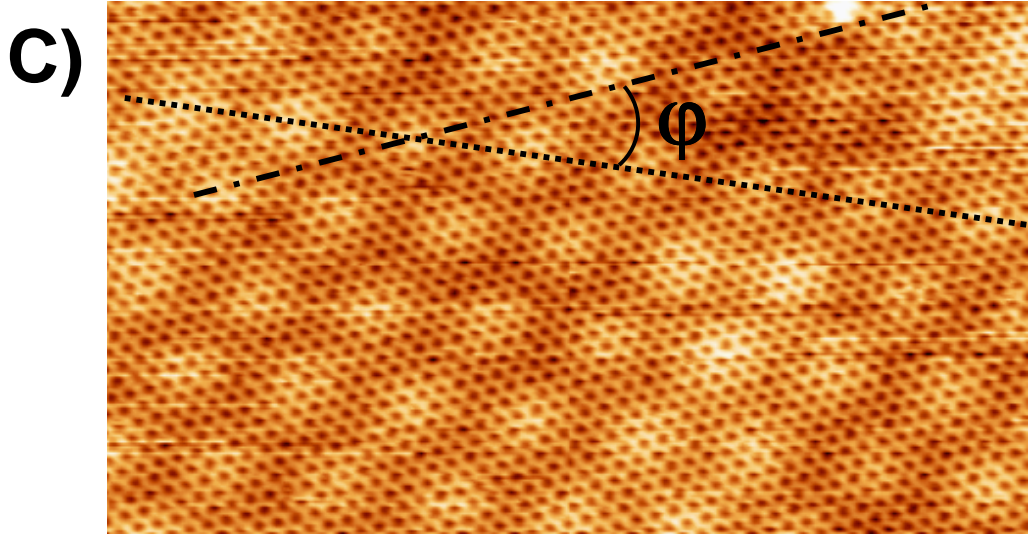

Figure 2-c

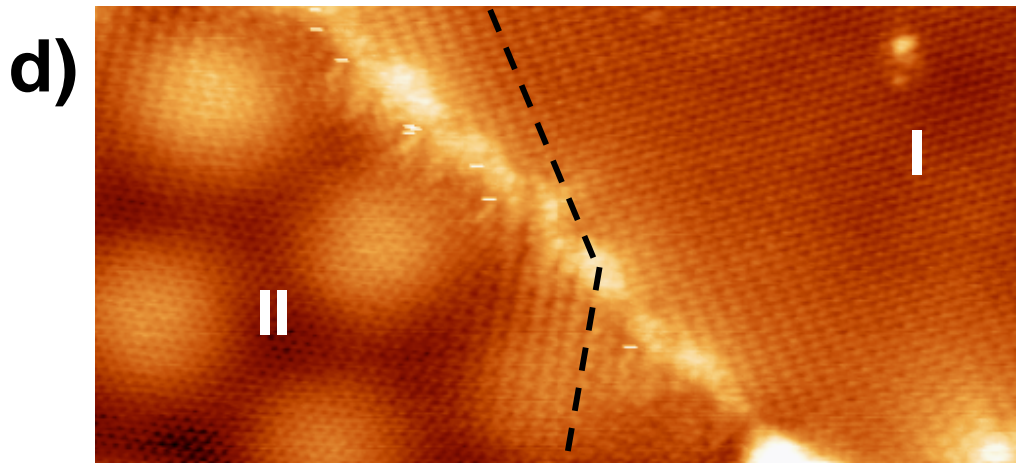

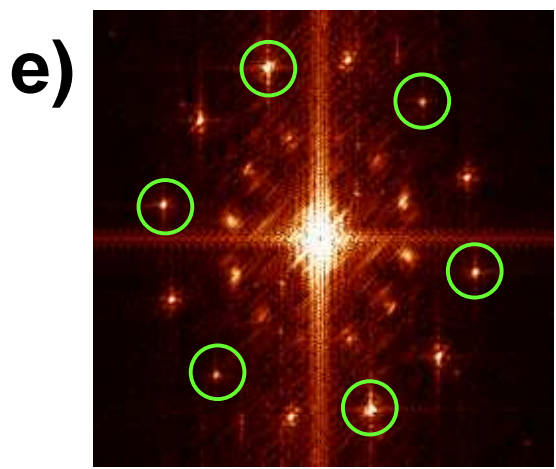

Figure 2 d-e.

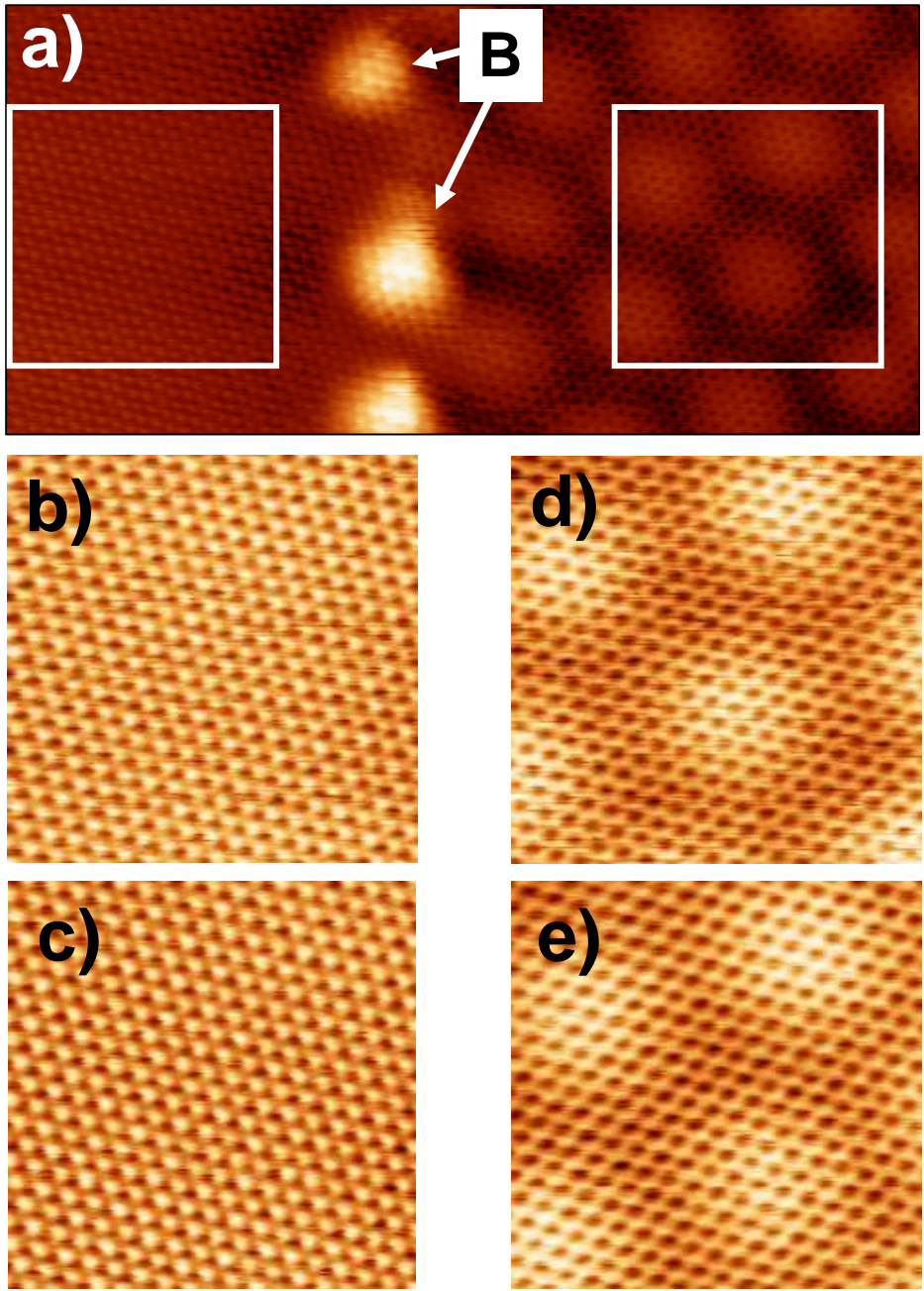

Figure 3